\documentstyle[aps,prd,epsf]{revtex}
\begin{document}
\draft
\title{Non-equilibrium effects on particle freeze-out in the early universe}
\author{Steen Hannestad}
\address{Theoretical Astrophysics Center, Institute of Physics and Astronomy,
University of Aarhus, 
DK-8000 \AA rhus C, Denmark}
\date{\today}
\maketitle

\begin{abstract}
We investigate the possible effects that deviations from kinetic
equilibrium can have on massive particles as they decouple from
chemical equilibrium. 
Different methods of solving the Boltzmann equation yield significantly
different relic number densities of such particles.
General considerations concerning the Dirac or Majorana
structure of the particles are discussed. It is shown that non-equilibrium
effects are small for particles decoupling while strongly 
non-relativistic, as will be the case for most cold dark matter 
candidates.
\\keywords: Early universe, dark matter, neutrino physics
\end{abstract}


\section{Introduction}
\label{sec:Intro}
Deviations from thermodynamic equilibrium are of fundamental importance
in the early universe. The examples are plentiful, for example the
freeze-out of photon thermodynamic equilibrium due to hydrogen recombination
\cite{recombination}
or the equivalent freeze-out of neutrinos slightly prior to 
electron-positron annihilation \cite{lighte}.
These phenomena have been investigated many times in the literature,
but almost always in the context of deviations from chemical
equilibrium, that is, kinetic or scattering equilibrium is assumed to
hold at all times. In the limit of Boltzmann statistics
the single particle distribution then takes the form
\begin{equation}
f(E) = e^{-(E-\mu)/T},
\end{equation}
where $\mu$ is a pseudo-chemical potential, independent of $E$.
In recent years there has been some discussions as to the validity of this
assumption, especially concerned with the decoupling of neutrinos
in the early universe
\cite{hannestad,DHS,kawasaki,dolgov1,dolgov2,kainu1,fields}. 
In this case it has been found by several
authors that the effects are small, but not negligible.
In the present paper we will discuss chemical versus scattering
equilibrium in a general context, where we limit ourselves to
generic expressions for the matrix elements.
Dolgov \cite{dolgov1} has provided a general discussion of deviations
from kinetic equilibrium caused by the overall expansion of the universe,
looking only at elastic scattering. The case we treat is where
both scattering and annihilation is taken into account
to solve the full Boltzmann equation, however
we resort to a numerical solution of the Boltzmann equation.

In general, for Dirac fermions it can be assumed that s-wave ($J=0$)
annihilation is by far the most important for non-relativistic
particles. The reason is the following: since particle and antiparticle
are two different particles there are no anti-symmetry requirements
and the annihilation can proceed via s-waves.
Since higher order terms $J \geq 1$ are suppressed by factors
of $v_{\rm rel}^{2J} \ll 1$ \cite{griest}, where $v_{\rm rel}$ is
the relative velocity of the incoming particles, 
s-wave annihilation is dominant. For Majorana particles
particle and antiparticle are identical objects and the combined
wave function must be antisymmetric. Therefore annihilation proceeds
primarily via p-waves ($J=1$), meaning that the annihilation amplitude
is suppressed relative to the Dirac case.
In the case of s-wave annihilation
 the matrix element is not momentum dependent
and the annihilation amplitude is constant. 
For p-wave annihilation there will, as mentioned, be a
dependence on the square of the 
relative velocity of the two incoming particles.
For non-relativistic particles we can therefore in general 
write the annihilation amplitude as
\begin{equation}
\sum |M|^2 = \cases{G^2&for J=0, \cr G^2 v_{\rm rel}^2 & for J=1. },
\end{equation}
where $G$ is a dimensionless coupling constant.

We shall use this expression for the matrix element even for relativistic
decoupling, although that is of course at best only an approximation
to the true matrix element. For example, in the case of massive Dirac
neutrinos, the matrix element will be strongly energy dependent
as long as the neutrinos are relativistic, but approach a constant as the
neutrinos become non-relativistic \cite{dolgov2}.
However, taking our approach, although quantitatively not accurate
for relativistic decoupling, still gives a very clear picture of
when non-equilibrium effects are important.

The elastic scattering part will, in general, have a momentum
dependent amplitude. This is for example the case for scattering
of heavy neutrinos on massless particles
 via the weak interaction. Here, the squared 
matrix element is proportional to the squared energy of the scatterer.
For the sake of simplicity we shall deal only with three cases, 
the first being
no elastic scattering at all and the second elastic scattering with a constant
amplitude. The first case will represent an extreme upper limit
to the size of non-equilibrium effects.
The second case, albeit not realistic, serves to indicate how elastic
scattering influences the shape of the massive particle spectra.
The third case we shall treat is the case where full scattering equilibrium
is maintained at all times.

\section{boltzmann equations}

When discussing non-equilibrium effects in the early universe the
relevant equation is the Boltzmann collision equation. Since we are
not discussing such issues as medium effects we shall use the
single-particle Boltzmann equation. 
Also, we restrict the calculation to using Boltzmann statistics, which
simplifies the calculations while leaving the essential physics intact.

Then the Boltzmann equation takes the form \cite{bernstein}
\begin{equation}
\partial f = C_{\rm ann}+C_{\rm el},
\label{eq:boltz}
\end{equation}
where
\begin{equation}
\partial f = \frac{\partial f}{\partial t} - H p \frac{\partial f}
{\partial p},
\end{equation}
$H$ being the Hubble parameter $H \equiv \dot{R}/R$.
On the right-hand side,
 $C_{\rm ann}$ and $C_{\rm el}$ represent annihilations and elastic
scatterings respectively.
These collision terms can be written generically as 
\begin{eqnarray}
C_{\text{coll}}[f] & = & \frac{1}{2E_{1}}\int d^{3}\tilde{p}_{2}
d^{3}\tilde{p}_{3}d^{3}\tilde{p}_{4}
(f_{3}f_{4}-f_{1}f_{2}) 
\label{integral}\\ 
& & \,\,\, \times S \sum | \! M \! |^{2}_{12\rightarrow 34}\delta^{4}
({\it p}_{1}+{\it p}_{2}-{\it p}_{3}-{\it p}_{4})(2\pi)^{4}, 
\nonumber
\end{eqnarray}
as long as we are only concerned with 2-body collisions.
Here we have $d^{3}\tilde{p}
 = d^{3}p/((2 \pi)^{3} 2 E)$. $S$ is a symmetrisation
factor of 1/2! for each pair of identical particles in initial or 
final states \cite{wagoner},
and $\sum \mid\!\!M\!\!\mid^{2}$ is the interaction matrix element
squared and spin-summed.
${\it p}_{i}$ is the four-momentum of particle $i$.
Now, the next issue is to actually perform the relevant 
collision integrals. It is in general possible to integrate $C_{\rm ann}$
and $C_{\rm el}$ down to two dimensions, depending on the actual shape
of $\sum |M|^2$. The details of these calculations are given in
the Appendix.

The usual approach to solving the above equation, Eq.~(\ref{eq:boltz}),
is to assume full scattering equilibrium (our third case). Then the
elastic scattering term is identically zero at all times, and
the Boltzmann can easily be integrated to yield \cite{kolb}
\begin{equation}
\dot{n}+3Hn = -\langle \sigma |v| \rangle (n^2-n_{\rm eq}^2),
\end{equation}
where $n$ is the actual number density and $n_{\rm eq}$ is the
equilibrium number density. $\langle \sigma |v| \rangle$ is 
the velocity-averaged annihilation cross-section \cite{GG91}.

As for the general expansion of the universe we shall assume that the
universe is completely radiation dominated and further we will ignore
entropy production in order to make the discussion completely generic.
Including the correct expansion law by using the Friedmann equation
and the equation of energy conservation will only perturb the general
results slightly in most cases.
Using the Friedmann equation \cite{kolb},
\begin{equation}
H^2 = \frac{8 \pi G \rho}{3},
\end{equation}
one then arrives at the following 
time-temperature relation
\begin{equation}
t(s) = 1.71 \, g_*^{-1/2} T_{\rm MeV}^{-2},
\end{equation}
where $g_* \equiv \rho/\rho_{\gamma}$, and $\rho$ is the total 
radiation energy density.

\section{numerical implementation}

In order to investigate the non-equilibrium 
effects in a quantitative way we still
have to calculate the remaining two dimensions of the collision
integrals, Eq.~(\ref{integral}),
 numerically. We have done this in order to estimate whether
non-equilibrium effects can ever be important during particle freeze-out.
Notice that we do not make any assumptions as to the size of the 
non-equilibrium
effects. That is, we do not assume {\it a priori} that they are small,
as was done for instance in Ref.~\cite{dolgov1},
but instead we perform a full numerical integration of the whole
system.
In all calculations we will assume that the massless particles are
in full thermodynamic equilibrium because of their interactions with
other light particles, an assumption which simplifies the calculations.
For both Dirac and Majorana particles one can expect that the
coupling constant will in general 
be the same for annihilation and scattering,
e.g.\ $G \propto G_F$ for weak interactions 
fFor a non-relativistic Dirac neutrino, $G = 32 m^4 G_F^2$,
assuming only the annihilation channel $\nu_H \bar{\nu}_H \to
\nu_e \bar{\nu}_e$) or
$G \propto \alpha$ for electromagnetic interactions. We can therefore scale
the amplitude in units of a single coupling constant, $G$, so that
\begin{equation}
\sum |M|^2 \propto G^2.
\end{equation}

\begin{figure}[h]
\begin{center}
\epsfysize=11truecm\epsfbox{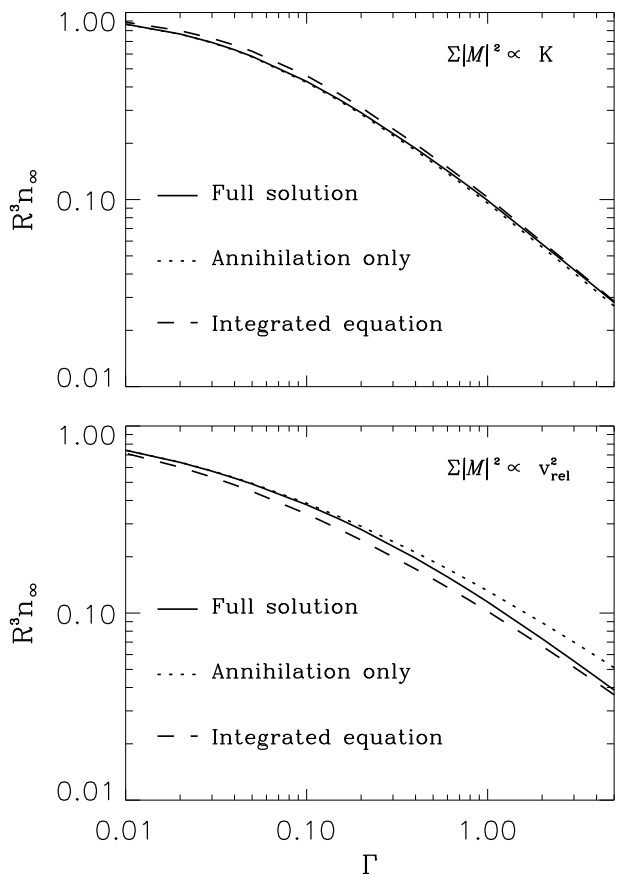}
\vspace{0truecm}
\end{center}
\vspace*{1cm}
\baselineskip 17pt
\caption{The final number density in comoving coordinates after
complete decoupling. The upper panel shows the case where the interaction
matrix element is constant, both for scattering and annihilation.
The lower panel shows the case for a velocity dependent annihilation
matrix element, still with a constant scattering matrix element.}

\label{fig1}
\end{figure}

Clearly, the size of the non-equilibrium effects will depend on $G$.
If $G$ is very small, particles will decouple while still relativistic.
This of course would not produce any non-equilibrium effects at all.
On the other hand, if the particles are extremely non-relativistic
before they decouple, the effects are also likely to be small
for the following reason. For very non-relativistic particles in 
equilibrium, the number density is suppressed exponentially relative to
the massless species. Therefore the annihilation
rate is also suppressed exponentially, while the scattering rate is
suppressed at most by some power of $p/m$. This will increase the
importance of elastic scattering relative to annihilation and lessen
any non-equilibrium effects, even though the scattering on massless
particles is essentially momentum conserving.
{\it A priori} one therefore expects that non-equilibrium effects are
the largest for semi-relativistic decoupling.

In Fig.~1 we have plotted the final number density of massive particles
in comoving coordinates as a function of a dimensionless
 interaction strength, defined as
\begin{equation}
\Gamma \equiv 5.20 \times 10^{20} \, m^{-1}_{\rm MeV} g_*^{-1/2} G^2.
\label{eq:gamma}
\end{equation}

There is a pronounced difference here between the two cases
of $J=0$ and $J=1$. 
For strong interactions (large $\Gamma$), the final number density 
is much higher for the velocity dependent interaction because of the
large suppression factor $v_{\rm rel}^2/m^2$. On the other hand, for 
relativistic decoupling, the velocity dependent interaction is stronger
by the same factor for the same value of $\Gamma$ and therefore the final
number density in this case is lower than for the constant matrix element.
Also, the freeze-out process proceeds much faster in the $J=1$ case
than in the $J=0$ because the annihilation amplitude falls off with
decreasing temperature.

\begin{figure}[h]
\begin{center}
\epsfysize=8truecm\epsfbox{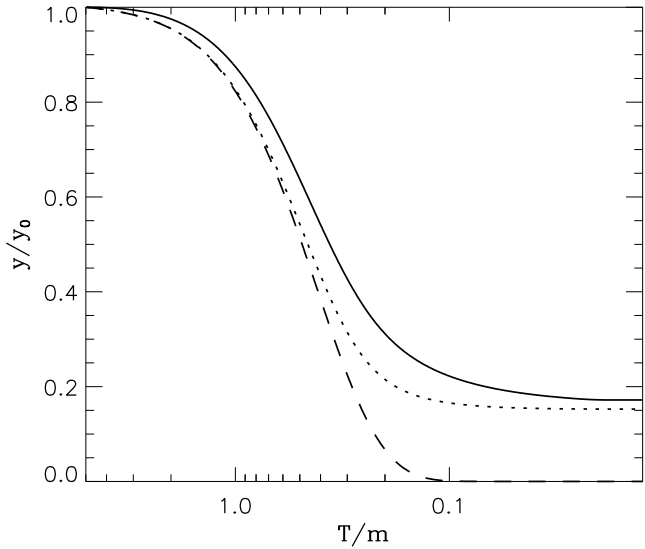}
\vspace{0truecm}
\end{center}
\baselineskip 17pt
\caption{Evolution of comoving number density $y \equiv nR^3$ 
in a specific case, where we have taken $\Gamma=0.5$.
The full curve shows the constant matrix element ($J=0$), the
dotted shows the velocity dependent matrix element ($J=1$), and the
dashed line is the equilibrium number density.}
\
\label{fig2}
\end{figure}

In Fig.~2 we show how the annihilation actually proceeds in the two
cases. This example is taken for $\Gamma = 0.5$. Clearly, for the
constant matrix-element decoupling is very prolonged, taking place over
a wide range in $T$. For the velocity dependent matrix element, decoupling
is much more sudden, essentially completing over a factor of two in $T$.
This difference will of course give rise to very different non-equilibrium
effects.

Fig.~3 shows the deviation in final number density between relative to 
that found by using the integrated Boltzmann equation.
For the solution of the full equation (the full lines)
We see the expected trend, namely that
relativistic decoupling leads to no significant non-equilibrium
effects and the same being true for very non-relativistic decoupling.
The effect has a maximum for semi-relativistic decoupling, where
the difference between using the full Boltzmann equation and the
integrated version
can be as large as 10-15\%. However, even for strongly
non-relativistic decoupling there still is a non-negligible effect
because of residual annihilations taking place after decoupling
from both scattering and chemical equilibrium.
Thus, one cannot entirely neglect non-equilibrium effects when doing
this sort of calculation.
Indeed, for the special case of massive neutrino decoupling it has been
demonstrated previously in the literature that out-of-equilibrium effects
can be significant \cite{hannestad,DHS,kawasaki,dolgov1,dolgov2,kainu1,fields}.
Note also the interesting feature that for the case of a constant matrix 
element, the annihilation term will itself preserve scattering equilibrium
if the massive particles are non-relativistic.
This can be seen directly from the kernel in the
annihilation integral, Eq.~(\ref{eq:kernel}).
The dotted curve in the upper panel of
Fig.~3 shows an example of this behaviour.
Here, the size of the non-equilibrium effects also drops for strong
interactions because complete equilibrium is maintained until the
particles are strongly non-relativistic, where after $C_{\rm ann}$
preserves scattering equilibrium independent of the interaction strength.
For the velocity dependent matrix element the situation is the reverse
because the annihilation integral kernel, Eq.~(\ref{eq:kernel2}),
 does not preserve kinetic
equilibrium. 
Rather, high momentum states annihilate much faster than
low momentum ones, so that kinetic equilibrium is quickly destroyed.
This can lead to very large errors in the computed number density
if one uses the approach of only taking into account the annihilation
term in the collision integral.
\begin{figure}[h]
\begin{center}
\epsfysize=11truecm\epsfbox{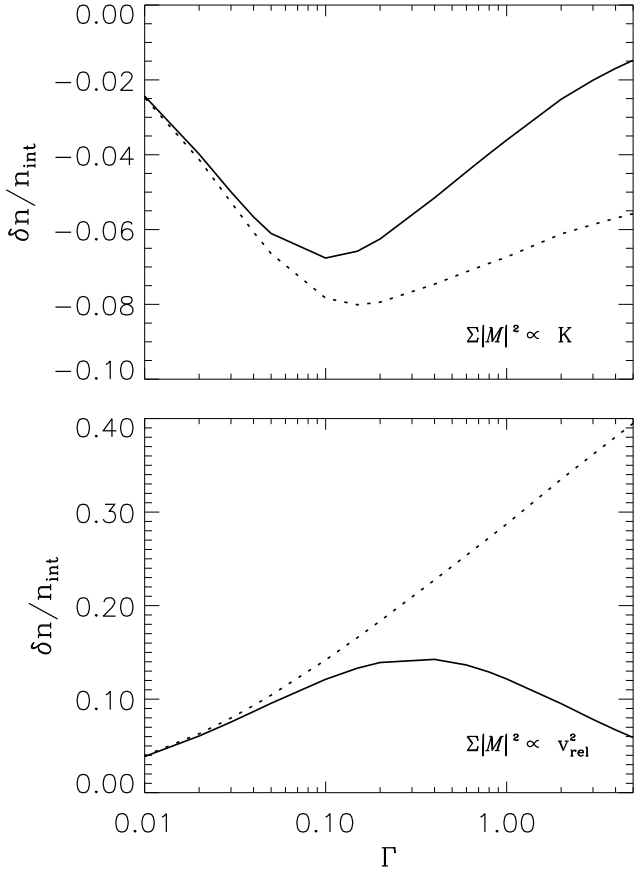}
\vspace{0truecm}
\end{center}
\vspace*{1cm}
\baselineskip 17pt
\caption{The deviation in number density depending on the calculational
approach. The full line is the deviation found found by using the
full Boltzmann equation relative to using the integrated equation.
The dotted line is the difference between using the annihilation
term only in the full Boltzmann equation and the integrated equation.}

\label{fig3}
\end{figure}

The sign of the non-equilibrium effect is also opposite for the $J=0$ and
$J=1$ scenarios. For the constant matrix element low momentum states will,
in general, annihilate faster than high momentum ones so that maintaining
kinetic equilibrium leads to a lower annihilation rate.
For the velocity dependent matrix element high momentum states deplete
quickly so that annihilation is slowed unless they are refilled by 
scattering. This effect is known from the case of massive Majorana
neutrinos \cite{hannestad,DHS,kawasaki}.

Finally, in Fig.~4 we show an example of the actual distribution
functions obtained by different methods, taken as a snapshot at
$T/m = 0.15$ for $\Gamma = 0.5$. 
What is shown is the ratio between the actual distribution function
and the corresponding equilibrium distribution with the same number
density.
Obviously, the distribution one obtains
by using the integrated equation is an equilibrium distribution per
default. However there can be huge differences between that and the actual
distribution function obtained from the full Boltzmann equation.
Also, there is a large difference between the full solution and the
one obtained by using only the annihilation term.
\begin{figure}[h]
\begin{center}
\epsfysize=11truecm\epsfbox{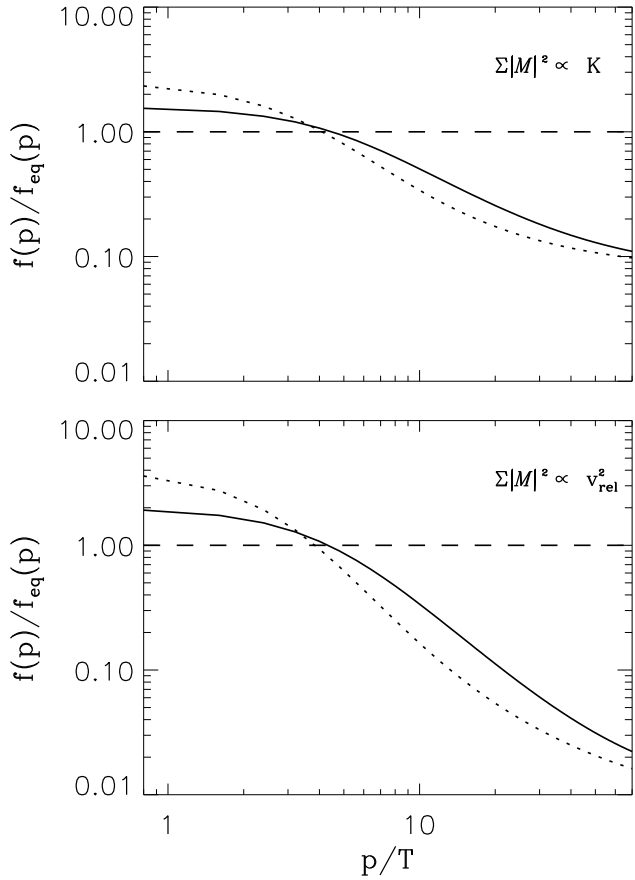}
\vspace{0truecm}
\end{center}
\vspace*{1cm}
\baselineskip 17pt
\caption{Actual distribution function of the heavy particle. The full line
corresponds to the solution of the full Boltzmann equation, the dotted
to including only annihilation and the dashed to the equilibrium
distribution in the integrated Boltzmann equation.
The distributions have been taken at a temperature of $T/m = 0.15$ and
for $\Gamma = 0.5$.}

\label{fig4}
\end{figure}

\section{discussion}

Our results clearly show that for the case of semi-relativistic decoupling
one cannot safely ignore the non-equilibrium effects pertaining to the
deviations from scattering equilibrium.
This will for instance be the case with MeV mass neutrinos which can 
influence Big Bang nucleosynthesis.

However, for either relativistic or non-relativistic decoupling the 
effects are negligible. Thus, for most cold dark matter candidates
which per definition decouple while strongly non-relativistic
\footnote{An example of a CDM candidate with radically different
properties is the axion which is produced in a condensate at
$T \gg m$, see for instance  Ref.~\cite{kolb}},
it is completely safe to assume full scattering equilibrium until long
after decoupling from chemical equilibrium.

As a simple example we shall consider a massive Dirac neutrino which is
thermally produced in the early universe. This particle is an
excellent CDM candidate if it has a mass in the 1 GeV region,
around the Lee-Weinberg limit \cite{LeeWein}.
The dimensionless interaction strength for this particle is given by
\begin{equation}
\Gamma_{\nu_D} \simeq 2.3 \, g_*^{-1/2} m_{\rm MeV}^3,
\end{equation}
assuming that the only relevant annihilation process is 
$\nu_H \bar{\nu}_H \to \nu_e \bar{\nu}_e$.
In Fig.~5 we have plotted $\Gamma$ as a function of the heavy neutrino mass.
For a mass
of 1 GeV, $\Gamma \simeq 7 \times 10^8 \ll 1$, meaning that there are
essentially no deviations from kinetic equilibrium.
As a note of caution we stress that we have assumed a scattering matrix
element of the same magnitude as the annihilation
matrix element which is incorrect. Rather, the scattering
matrix element is $\sum |M|^2_{\rm scattering} \simeq G_F^2 m^2 T^2$
which is lower by a factor of $T^2/m^2$. Even so there will be no
discernible effect due to deviations from scattering equilibrium for
such a heavy neutrino.
The same will be true for super-symmetric dark matter particles which
are likely to have a mass in the multi-GeV region or above \cite{SSDM}.
However, there are other examples where non-equilibrium effects
are sizeable, for example the previously mentioned example of an MeV
neutrino.
\begin{figure}[h]
\begin{center}
\epsfysize=8truecm\epsfbox{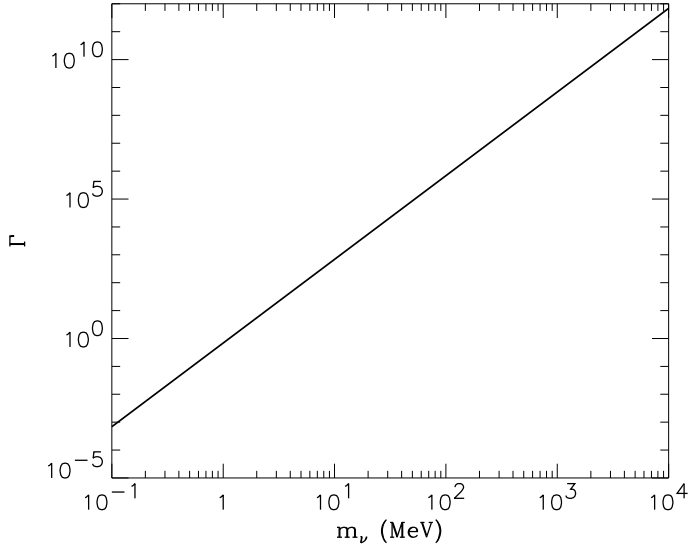}
\vspace{0truecm}
\end{center}
\vspace*{1cm}
\baselineskip 17pt
\caption{The dimensionless interaction strength $\Gamma$, defined in
Eq.~(\ref{eq:gamma}), for a non-relativisitc Dirac neutrino as a
function of the mass.}

\label{fig5}
\end{figure}

It was also shown that taking into
account only the annihilation term in the full Boltzmann equation
can lead to large errors in the final number density if the matrix
element is velocity dependent. This was for instance the case in the
work by Kawasaki et al. \cite{kawasaki}.

Finally, in the present paper we have not discussed possible deviations from
equilibrium in the massless particle species induced by interactions
with the massive particles. This effect is for instance of importance
when the massless particles themselves are decoupling from equilibrium
at the same temperature as the massive ones. An example is the decoupling
of MeV $\tau$-neutrinos where the induced deviations from equilibrium in
the electron neutrino distribution causes Big Bang Nucleosynthesis
to change \cite{hannestad,DHS,dolgov2,kainu1,fields}.

\acknowledgements
This work was supported by the Theoretical
Astrophysics Center under the Danish National Research Foundation.

\appendix
\section*{collision integrals}

This appendix deals with the phase-space integrations of the collision
operator of Eq.\ (\ref{integral}). The principal method of integration
was developed in Ref.~\cite{heating} and in the following we only
give a brief outline and further develop the integrations in the 
non-relativistic case.

The collision integral in the general two-body collision integral can
be reduced to two dimensions. Let the process be 
\begin{equation}
1 + 2 \to 3 + 4.
\end{equation}
Then the result is
\begin{equation}
C_{\rm coll}[f_1(p_1)] = \int dp_2 dp_3 
[J(p_1,p_2,p_3)]^{\inf[1,\alpha^+]}_{\sup[-1,\alpha^-]}
(f_3 f_4 - f_1 f_2),
\end{equation}
where 
\begin{eqnarray}
\alpha^\pm & = & \frac{-2 \gamma - 2 p_{2}^{2} - Q
\pm 2 p_{2} (2 \gamma + p_{1}^{2} + p_{2}^{2} + p_{3}^{2} + Q)
^{\frac{1}{2}}}{2 p_{1}p_{3}}, \\
\gamma & = & E_{1} E_{2} - E_{1} E_{3} - E_{2} E_{3}, \\
Q & = & m_{1}^{2} + m_{2}^{2} + m_{3}^{2} - m_{4}^{2}.
\end{eqnarray}
The function $J(p_1,p_2,p_3)$ is the matrix element integrated over all
variables except $p_2$ and $p_3$ \cite{heating}.

In the non-relativistic case the expression for $\alpha$ simplifies 
further, and it becomes possible to do one more integration.
In this case on gets 
\begin{equation}
\alpha^\pm =  \cases{\frac{2m(p_3-m)\pm p_2(2m-p_3)}
{p_1 p_3}&for annihilation, \cr 
\frac{p_1^2+p_3^2-2p_2^2 \pm 2 p_2^2}
{2 p_1 p_3} & for scattering. }
\end{equation}
In the case where the light particles are in equilibrium we can 
further integrate over either $dp_3$ (annihilation) or $dp_2$
(scattering).  
We shall treat the case of annihilation first. For the case
of a constant matrix element one gets the result
\begin{eqnarray}
I  \equiv  \int_{p_{3,{\rm min}}}^{p_{3,{\rm max}}} 
[J(p_1,p_2,p_3)]^{\inf[1,\alpha^+]}_{\sup[-1,\alpha^-]}
(f_3 f_4 - f_1 f_2)& = &
G^2 \frac{1}{2} F_1 (p_1,p_2),
\label{eq:kernel}
\end{eqnarray}
where
\begin{equation}
F_1 (p_1,p_2) = -f(p_1)f(p_2)+e^{-(E_1+E_2)/T}.
\end{equation}
For a matrix element that is proportional to $v_{\rm rel}^2$,
one gets instead
\begin{equation}
\int_{p_{3,{\rm min}}}^{p_{3,{\rm max}}} 
[J(p_1,p_2,p_3)]^{\inf[1,\alpha^+]}_{\sup[-1,\alpha^-]}
(f_3 f_4 - f_1 f_2) = \frac{p_1^2+p_2^2}{m^2} I.
\label{eq:kernel2}
\end{equation}
Using these expressions we have reduced the pair-annihilation integral
to a one-dimensional integral which is quite straightforward to
calculate numerically,
\begin{eqnarray}
C_{\rm ann} & = & \frac{1}{(2\pi)^3} \frac{1}{2E_1}
\int \frac{p_2^2 dp_2}{2E_2} I F_1 (p_1,p_2).
\end{eqnarray}
Also, it is easy to see that the velocity
dependent matrix element is suppressed relative to the constant one
by a factor $p^2/m^2$, which is large in the non-relativistic
case.
From the above equations one can also easily calculate the velocity
averaged annihilation cross sections for the two cases if one assumes
scattering equilibrium at all times
\begin{equation}
\langle \sigma |v| \rangle =  G^2 \times 
\cases{\frac{1}{32 \pi m^2}&for $J=0$, \cr 
\frac{3T}{16 \pi m^3} & for $J=1$. }
\end{equation}
These can be plugged into the standard integrated Boltzmann equation
\begin{equation}
\dot{n}+3Hn = -\langle \sigma |v| \rangle (n^2-n_{\rm eq}^2).
\end{equation}

Finally, the evaluation of the scattering integrals in the 
non-relativistic limit is also quite straightforward.
Here, the integral must be over $dp_2$ to avoid integrating over the
distribution function of the heavy particle. 
The result of this integration is
\begin{eqnarray}
H \equiv \int_{p_{2,{\rm min}}}^{p_{2,{\rm max}}} 
[J(p_1,p_2,p_3)]^{\inf[1,\alpha^+]}_{\sup[-1,\alpha^-]} 
(f_3 f_4 - f_1 f_2)& = &
G^2 \frac{2}{p_1 p_3}\left[\exp\left(-\frac{|p_1-p_3|}{2}\right)
\right. \nonumber \\
&& \left. -\exp\left(-\frac{p_1+p_3}{2}\right)\right] F_2 (p_1,p_3),
\end{eqnarray}
where
\begin{equation}
F_2 (p_1,p_3) = -f(p_1)+f(p_3) \, e^{-E_1-E_3}.
\end{equation}
Altogether then we can write the elastic scattering term as
\begin{eqnarray}
C_{\rm el} & = & \frac{1}{(2\pi)^3} \frac{1}{2E_1}
\int \frac{p_3^2 dp_3}{2E_3} H F_2 (p_1,p_3).
\end{eqnarray}

\end{document}